\documentclass[12pt,twoside,pointlessnumbers,nofootinbib,smallheadings]{revtex4}
\usepackage{amsmath}
\usepackage{amssymb}
\usepackage{color}
\usepackage{makeidx}

\usepackage{hangcaption,fancybox,feynmp}
\usepackage{indentfirst}

\usepackage{graphicx}
\usepackage{epsfig,subfigure,psfrag}
\usepackage{CJK}
\usepackage{mathrsfs}

\def \be {\begin{equation}}
\def \ee {\end{equation}}
\def \ba {\begin{eqnarray}}
\def \ea {\end{eqnarray}}
\newcommand {\dbar}{{d\kern-.22em\lower-.73ex\hbox{-}}}

\newcommand {\fslash}[1]{{#1\kern -0.6em / \kern 0.2em}}

\makeindex

\begin{document}

\title{
Complex Vacuum and Lightness of Higgs Boson
}
\author{ Shou-hua
Zhu  }

\affiliation{
$ ^1$ Institute of Theoretical Physics $\&$ State Key Laboratory of
Nuclear Physics and Technology, Peking University, Beijing 100871,
China \\
$ ^2$ Center for High Energy Physics, Peking University,
Beijing 100871, China \\
$ ^3$ Collaborative Innovation Center of Quantum Matter, Beijing, China
}

\date{\today}

\maketitle

\begin{center}
{\bf Abstract}
\end{center}

The CP violation observed in K- and B-mesons can be successfully accommodated via the CKM matrix in the standard model. However the additional CP violation is required in order to induce the matter and anti-matter asymmetry observed in the Universe. The additional CP violation can be induced by complex vacuum, namely the spontaneous CP violation.
 In this paper we reveal the intimate connection of the spontaneous CP violation and the lightness of the Higgs boson, which was recently discovered by ATLAS and CMS at the LHC.




\newpage



\section{Introduction}

Recently ATLAS \cite{:2012gk} and CMS  \cite{:2012gu} at Large Hadron Collider have discovered one new particle around 125 GeV with compatible properties with Higgs boson in the standard model (SM) of high energy physics. In the SM, the mass of Higgs boson is a free parameter and could be enormous.
The great success of the {\em renormalizable} SM indicates that there should exist large hierarchy between electro-weak and the higher scale, based on the renormalization group flow
analysis of K. Wilson. The Higgs boson, as the spin-0 scalar, tends to get contributions from the higher scale and becomes enormous.
Therefore the pursuit of underlying reason why the Higgs boson is light is quite natural.
Even before the direct confirmation of the light Higgs boson, one has inferred the similar Higgs boson information from the precision measurements at LEP and Tevatron. Moreover the new ideas have been proposed in order to account for the lightness of the Higgs boson. For example, the lightness of Higgs boson could be protected by supersymmetry. In the minimal supersymmetric model (MSSM), the mass of Higgs boson can be expressed as $m_H < m_Z |\cos2\beta|$ at tree level, where $m_Z$ is the Z boson mass and $\beta$ is the angle defined by $\tan\beta=v_2/v_1$. Here $v_2$ and $v_1$ are the vacuum expectation values (VEV) of the two Higgs fields introduced in MSSM. In the little Higgs models, the lightness of Higgs boson is protected by the approximate global symmetry and is treated as the pseudo-Goldstone boson. Though the supersymmtry and little Higgs models are quite intriguing, the experimental search for their companions, namely the gluinos and squarks in supersymmetric models and extra gauge bosons $W^\prime$ and/or $Z^\prime$ in little Higgs models, has failed. In this paper we will reveal another interesting connection between lightness of Higgs boson and the CP spontaneous violation.

CP violation is required in order to account for the matter and anti-matter asymmetry in the Universe, as pointed by Sakharov \cite{Sakharov:1967dj} several decades ago. In the SM, CP non-conservation does exist in the CKM matrix. However such CP violation is not sufficient to accommodate the observed asymmetry \cite{Riotto:1999yt}.
Seeking additional source of CP violation is one of the most important motivations of B-factories, super B-factories and LHCb.
In the SM the origin of CP violation in CKM matrix can be traced back to the complex Yukawa interaction among Higgs field and fermions. Besides such explicit CP violation,
CP non-conservation can also be traced back to the complex vacuum, as pointed by T.D. Lee \cite{Lee:1973iz,Lee:1974jb} long time ago.
Now that the required additional CP violation and Higgs boson obtain mass via the same mechanism, i.e. the spontaneous symmetry breaking, is the Higgs boson mass intimately related to such CP violation?
In this paper we will show that, taking a simplest spontaneous CP violation model as the example, the mass of the lightest neutral Higgs boson is closely related to CP violation phase of the vacuum.
Provided that the SM is not CP-violated, the lightest neutral Higgs boson would be
massless, sharing the same characteristic with the neutral Goldstone boson.
In reality we require additional CP violation from complex vacuum, as a consequence, the neutral lightest Higgs boson becomes massive and the other neutral
Higgs bosons are much heavier than the lightest one.
In this sense we relate the lightness of Higgs boson to the complex vacuum.
For the most general spontaneous CP violation models, we argue that the lightest neutral Higgs boson can have the exact the same feature as in the simplest case.

\section{Spontaneous CP violation and lightness of Higgs boson in two Higgs doublets model}

One decade ago, we have investigated the simplest spontaneous CP violation model \cite{Huang:1999xa,Huang:2001me} and applied it to the B-meson decay, especially
in the large $\tan\beta$ case. The motivations of this model are as following. If one insists the natural flavor conservation (NFC) condition,
a minimum of three Higgs doublets are necessary in order to have spontaneous CP violation \cite{Weinberg:1976hu,Branco:1980sz,Shizuya:1980jx}.
Provided that NFC is given up, the CP can be explicitly violated in the Higgs potential. However we are only interested in the spontaneous
CP violation and investigate its connection with Higgs boson mass. In this paper we will investigate the properties of the Higgs potential for
the simplest case of spontaneous
CP violation, i.e. only introducing two complex Higgs doublets without obeying NFC condition. In the previous works, we have introduced additional terms which break
NFC condition only softly \cite{Huang:1999xa,Huang:2001me}.

For two complex Higgs doublets $\Phi_i$ (i=1,2) with hypercharge $Y=1$,
$$
\Phi_i = \begin{pmatrix} \phi_i^+ \\ \phi_i^0 \end{pmatrix}; \Phi_i^\dagger = \begin{pmatrix} \phi_i^- \\ \phi_i^{0*} \end{pmatrix}
$$
$\phi^0_i$ and $\phi^\pm_i$ represent neutral and charged complex component of $\Phi_i$,
the most general CP-invariant Higgs potential can be written as
$$
V\left(\Phi_1, \Phi_2\right) =V_2+V_4.
$$

It'd better to divide $\Phi_i^\dagger \Phi_j$ term as real and imaginary parts separately: $Re[\Phi_i^\dagger \Phi_j]\equiv Re_{ij}$ and $Im[\Phi_i^\dagger \Phi_j]\equiv Im_{ij}$
in order to classify the terms conveniently.
For $i= j$ imaginary part is 0. Under CP transformation, only imaginary part for $i\neq j$ changes sign.
Hermitian and CP invariant require \footnote{ $m_{2}^2 Re_{12}$ is {\em not} the physical term since it can be rotated away by redefinition of fields and other parameters.}
\ba
V_2&=&  m_{1}^2 Re_{11}  + m_{3}^2 Re_{22}
\ea
with $m_i^2$ (i=1-3) the real parameters.

Hermitian and CP invariant also require
\ba V_4&=&Re_{11} \left[ \lambda_{1} Re_{11} +  \lambda_{2} Re_{12}  + \lambda_{3} Re_{22}  \right] \nonumber \\
        && + Re_{12} \left[ \lambda_{4} Re_{12}+ \lambda_{5} Re_{22} \right] \nonumber \\
             &&+  \lambda_{6} Re_{22}^2  + \lambda_{7} Im_{12}^2
\ea
where
$\lambda_i$ (i=1-7) are real parameters.

We assume that the minimum of the potential is at
\ba
<\Phi_1> =\begin{pmatrix} 0 \\ v_1 \end{pmatrix};
<\Phi_2> =\begin{pmatrix} 0 \\ v_2 e^{i\xi} \end{pmatrix};
\ea
which breaks the gauge group $SU(2)_L \otimes U(1)_Y$ down to $U(1)_{em}$ and the CP invariance.
We can define $\tan\beta=v_2/v_1, v^2=v_1^2+v_2^2$. In the end the $v$ is determined by the mass of weak gauge bosons, as usual.

The requirement of the stationary point
of the potential leads to the following constraints:
\ba
\frac{\partial{V}}{\partial v_1} &=& 2 v_1 \left[ m_1^2+2 \lambda_1 v_1^2+  \lambda_3  v_2^2 \right]+\left[ 3 \lambda_2 v_1^2+
\lambda_5 v_2^2\right]v_2 \cos\xi \nonumber \\
&& + 2 \lambda_4 v_1 v_2^2 \cos^2\xi + 2 \lambda_7 v_1 v_2^2 \sin^2\xi = 0
\ea

\ba
\frac{\partial{V}}{\partial v_2} &=& 2 v_2 \left[ m_3^2+2 \lambda_6 v_2^2+  \lambda_3  v_1^2 \right]+\left[ 3 \lambda_5 v_2^2+
\lambda_2 v_1^2\right]v_1 \cos\xi \nonumber \\
&& + 2 \lambda_4 v_1^2 v_2 \cos^2\xi + 2 \lambda_7 v_1^2 v_2 \sin^2\xi = 0
\ea

\ba
\frac{\partial{V}}{\partial \xi} &=& \sin\xi v_1 v_2 \left[2(\lambda_7-\lambda_4) v_1 v_2 \cos\xi -\lambda_2 v_1^2-\lambda_5 v_2^2 \right]=0 \label{zeroconorigin}
\ea

For $\sin\xi, v_1,v_2 \neq 0$,
we can solve the equations to get
\ba
0 &= &  2(\lambda_7-\lambda_4) v_1 v_2 \cos\xi-\lambda_2 v_1^2-\lambda_5 v_2^2
\ea
\ba
m_1^2 &= & -\left[  2\lambda_1 v_1^2+ \left( \lambda_3 + \lambda_7 \right) v_2^2  +  \lambda_2 v_1 v_2  \cos \xi \right]
\ea
\ba
m_3^2 &= & -\left[  2\lambda_6 v_2^2+ \left( \lambda_3 + \lambda_7 \right) v_1^2  +  \lambda_5 v_1 v_2  \cos \xi \right] \label{zerocon}
\ea
After substituting those conditions into Higgs potential, we can trade $v_1, v_2, \xi$ with $m_1^2,m_3^3, \lambda_2$.
It should be emphasized that $\xi=0$
is the trivial solution of the stationary conditions!
In this case Eq. \ref{zerocon} does not necessarily hold (cf. Eq.\ref{zeroconorigin}), the would-be-eliminated parameter by this equation will
eventually control the mass of pseudo-scalar, usually denoted as $A$ in literature.

The potential minimum conditions require that at the stationary point
\ba &&  \frac{\partial^2 V}{\partial v_1 \partial v_1} >0 \nonumber \\
 &&  \det \left(\frac{\partial^2 V}{\partial x_i \partial x_j}\right) >0 \nonumber \\
&&  \det \left(\frac{\partial^2 V}{\partial y_i \partial y_j} \right) >0\label{ieq}
\ea
where $x_i=v_1,v_2$  and $y_i=v_1,v_2,\xi$.

We show here explicitly the first inequality
\ba
\cos^2\beta \left( 4 \lambda_1 -3 \cos^2\xi  (\lambda_4-\lambda_7)\tan^2\beta-2 \lambda_5 \cos \xi)\tan^3\beta  \right) &>&0  \label{ieq1}
\ea
and the other two tedious inequalities can be easily deduced.

Now we switch to the spectrum of the physical Higgs bosons. The charged part can be written as $(\phi_1^-, \phi_2^-) M (\phi_1^+, \phi_2^+)$
where the mass matrix
\ba
M = -\lambda_7 \begin{pmatrix} v_2^2 & -v_1 v_2 e^{-i \xi} \\
-v_1 v_2 e^{i \xi} & v_1^2 \end{pmatrix}.
\ea

The mass eigenstates of $G^-$ and $H^- $ can be written as
\ba G^- &=& e^{i\xi} \sin\beta \phi_2^- + \cos\beta \phi_1^- \\
H^- &=& e^{i\xi} \cos\beta \phi_2^- - \sin\beta \phi_1^-
\ea
with
\ba
m_{G^\pm}=0; &&
m_{H^\pm}= -\lambda_7 v^2.
\ea
As usual the charged Goldstone boson $G^\pm$ will be absorbed by charged gauge bosons.

For the neutral part, in the basis of $\{Im (\phi_1), Im (\phi_2), Re(\phi_1), Re(\phi_2)\}$ the symmetric mass matrix $M$ can be expressed as
\ba
M_{11} &=& (\lambda_4-\lambda_7) v_2^2 \sin^2 \xi \nonumber \\ \nonumber
M_{12} &=& \lambda_5 v_2^2 \sin^2 \xi \\ \nonumber
M_{13} &=& \left[-\lambda_5 \tan\beta+ (\lambda_7-\lambda_4) \cos\xi \right] v_2^2 \sin \xi \\ \nonumber
M_{14} &=& \left[(\lambda_4-\lambda_7) v_1+ \lambda_5 v_2 \cos\xi \right] v_2 \sin \xi \\ \nonumber
M_{22} &=& 4 \lambda_6 v_2^2 \sin^2 \xi \\ \nonumber
M_{23} &=&\left[2 (\lambda_3+\lambda_7) v_1 + \lambda_5 v_2 \cos\xi \right] v_2 \sin \xi \\ \nonumber
M_{24} &=&\left[ \lambda_5 v_1 + 4 \lambda_6 v_2 \cos\xi \right] v_2 \sin \xi \\ \nonumber
M_{33} &=&
\frac{1}{2 v_1} \left[-4 \lambda_5 v_2^3 \cos\xi+ v_1 \left( 8 \lambda_1 v_1^2- 3 v_2^2 (\lambda_4-\lambda_7) (1+ \cos(2\xi) ) \right) \right] \\ \nonumber
M_{34} &=& v_2 \left[ (2\lambda_3 - \lambda_4+3 \lambda_7)v_1 \cos\xi -v_2 \lambda_5 \sin^2\xi \right] \\ \nonumber
M_{44} &=& (\lambda_4-\lambda_7) v_1^2+ 2 \lambda_5 v_1 v_2 \cos\xi+  2 \lambda_6 v_2^2 (1+\cos(2\xi)). \label{mmatrix}
\ea

From mass matrix $M$, we should note one important feature in the limit of $\xi \rightarrow 0$. In this limit, the whole mass matrix can be decomposed into
zero  and a non-zero $M^\prime_{2\times 2}$ matrices as
\ba
M= \begin{pmatrix} 0_{2\times 2} & 0_{2\times 2} \\
 0_{2\times 2} & M^\prime_{2\times 2}
 \end{pmatrix}.
 \ea
This feature can be understood because we have applied the constraint of Eq. \ref{zerocon}, which is not required for the case of CP conserving $\xi=0$. One less free
parameter drives one of physical neutral Higgs boson massless, sharing the similar feature with neutral Goldstone boson $G^0$.
This point can be shown by
rotating away Goldstone state $G^0 = \cos\beta Im \phi_1^0 + \sin\beta \cos\xi Im \phi_2^0-\sin\beta \sin\xi  Re (\phi_2^0) $.
We can obtain mass matrix $N_{3\times 3} v^2$ in the basis of \ba
\begin{pmatrix} -\sin\beta  Im (\phi_1^0) + \cos\beta \cos\xi Im (\phi_2^0)-\cos\beta \sin\xi Re (\phi_2^0) \\ Re (\phi_1^0) \\
\sin\xi Im (\phi_2^0)+ \cos\xi Re (\phi_2^0)   \end{pmatrix}.
\ea
Here the matrix $N$ can be expressed as
\ba
N_{11} &=& \left[\lambda_4-\lambda_7 \right]  \sin^2 \xi \nonumber \\ \nonumber
N_{12} &=& \sin\beta \left[\lambda_5 \tan\beta+ (\lambda_4-\lambda_7)\cos\xi \right]  \sin \xi \\ \nonumber
N_{13} &=& - \left[\lambda_5 \sin\beta+ (\lambda_4-\lambda_7)\cos\beta\cos\xi \right]  \sin \xi \\ \nonumber
N_{22} &=& \frac{1}{2}\cos^2\beta \left[ 8 \lambda_1-2 \cos\xi \tan^2\beta
\left( 3 (\lambda_4-\lambda_7)\cos\xi+2 \lambda_5 \tan\beta \right) \right]
  \\ \nonumber
N_{23} &=&
\frac{1}{2} \cos\beta\sin\beta\left[ 4 \lambda_3- \lambda_4+ 5 \lambda_7
-(\lambda_4-\lambda_7)\cos (2\xi)\right] \\ \nonumber
N_{33} &=&
\frac{1}{2}\left[2 (\lambda_4-\lambda_7)\cos^2\beta \cos^2\xi+4  \lambda_5\cos\beta\sin\beta \cos\xi +8 \lambda_6 \sin^2\beta \right].
\label{nmatrix}
\ea
In the limit of $\xi \rightarrow 0$, we can expand the determinant of mass matrix as
\ba
\det \left(Nv^2\right) &=&
2 \xi^2 \sin^2\beta v^6 \left[
\left( - 2 \lambda_1 \lambda_5^2+ 2 (\lambda_4-\lambda_7)(4 \lambda_1 \lambda_6-
(\lambda_3+\lambda_7 )^2 ) \right)\cos^2\beta  \right. \nonumber \\
&& -\tan\beta \left(2 \lambda_4
-2 \lambda_7+ \lambda_5\tan\beta \right) \left[ \lambda_5(\lambda_3+\lambda_6+\lambda_7 ) \right. \nonumber \\
&& \left. \left.+  \lambda_5(\lambda_3-\lambda_6+\lambda_7 )\cos (2\beta)+
2\lambda_6 (\lambda_4-\lambda_7) \sin(2\beta) \right] \right] +O(\xi^3). \ea
Trace of the mass matrix can be expanded around $\xi=0$ as
\ba
Tr\left(Nv^2\right) &=& v^2\sec\beta \left[ (3 \lambda_1+\lambda_6) \cos\beta+
( \lambda_1+\lambda_4-\lambda_6-\lambda_7)\cos(3\beta) + \lambda_5
(\sin(3\beta)-\sin\beta) \right] \nonumber \\
&& + v^2 \left[-\lambda_5 \cos(2 \beta) + 2 (\lambda_4 - \lambda_7) \sin(2 \beta) \right] \tan\beta \xi^2 +O(\xi^3). \ea

The mass of the lightest
neutral Higgs boson mass can be written, around $\xi=0$, as
 \ba
 m_{h_1} = f(\lambda_i, \beta) \xi^2 v.
 \ea
The mass eigenstate of this neutral Higgs boson approaches $ -\sin\beta  Im (\phi_1^0) + \cos\beta Im (\phi_2^0)$, orthogonal to $G^0$.
Compared with the mass of the SM Higgs boson which is solely determined by $\lambda$ and VEV, the lightest neutral Higgs boson here is also
determined by the CP violation parameter $\xi$. In the limit of $\xi \rightarrow 0$, the lightest neutral Higgs boson and the neutral Goldstone boson both are the mixing
states of $ Im (\phi_1^0)$ and $Im (\phi_2^0)$, sharing the same massless feature. Provided that the SM is not CP-violated, the lightest neutral Higgs boson would be
massless. The reality is that the SM {\em is} CP-violated \footnote{Here we refer to the additional CP violation. In fact, the CKM CP violation phase can also be traced back to the complex vacuum.}, therefore the lightest neutral Higgs boson is massive. In this sense, the lightness of the lightest Higgs boson
is intimately connected to the spontaneous CP violation.

The next question is: Does the mass of the physical lightest neutral Higgs boson in the most general spontaneous CP violation models also approach 0 in the CP invariant limit? The answer is {\em yes} if the $\xi=0$ is the trivial solution to equations of the stationary conditions! The arguments are
as following. Provided that the original
Higgs potential is CP-conserving with certain symmetry. In order to realize the spontaneous symmetry breaking, which gives mass to gauge bosons and induces CP violation, the stationary condition for $\xi$ will eliminate one free parameter in the Higgs potential as in Eq. \ref{zeroconorigin}.  If $\xi = 0$ is the trivial solution, the would-be-eliminated parameter must be associated with $\xi$ in the mass matrix (c.f. \ref{mmatrix} or \ref{nmatrix}). In the limit of
$\xi \rightarrow 0$, at least one neutral Higgs boson shares the massless nature of neutral Goldstone boson. Such behavior indicates that the lightness of the Higgs boson can be traced back to the spontaneous CP violation phase of the vacuum.

In the following, we show explicitly the neutral Higgs mass spectrum, especially the lightest one.
One can diagonalize the $3\times 3$ real symmetric mass matrix via
\ba V_n . N v^2 . V_n^T = \frac{1}{2} diag(m_{h_3}^2,m_{h_2}^2,m_{h_1}^2) \ea
assuming $m_{h_1} \leq  m_{h_2} \leq m_{h_3}$.
\ba
\begin{pmatrix} h_3 \\ h_2 \\
h_1   \end{pmatrix}= V_n^T\begin{pmatrix} -\sin\beta  Im (\phi_1^0) + \cos\beta \cos\xi Im (\phi_2^0)-\cos\beta \sin\xi Re (\phi_2^0) \\ Re (\phi_1^0) \\
\sin\xi Im (\phi_2^0)+ \cos\xi Re (\phi_2^0)   \end{pmatrix}
\ea

\begin{figure}[htbp]
 \begin{center}
  \includegraphics[width=0.8\textwidth]{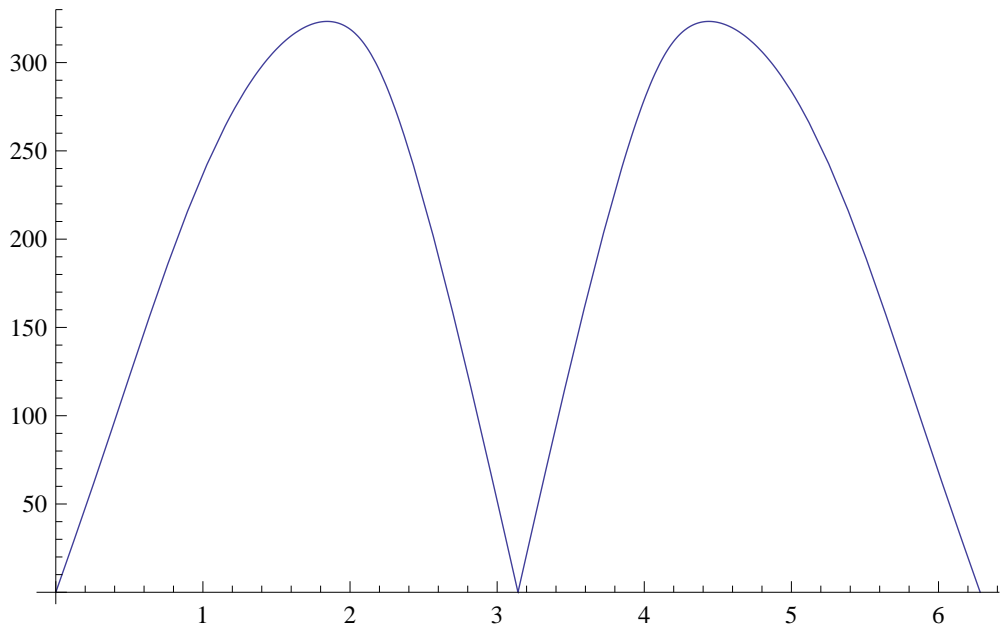}
  \end{center}
  \caption{ Lightest Higgs boson  mass  $m_{h_1}$[in GeV] as a function of $\xi$. }
  \label{h1}
\end{figure}

\begin{figure}[htbp]
 \begin{center}
  \includegraphics[width=0.8\textwidth]{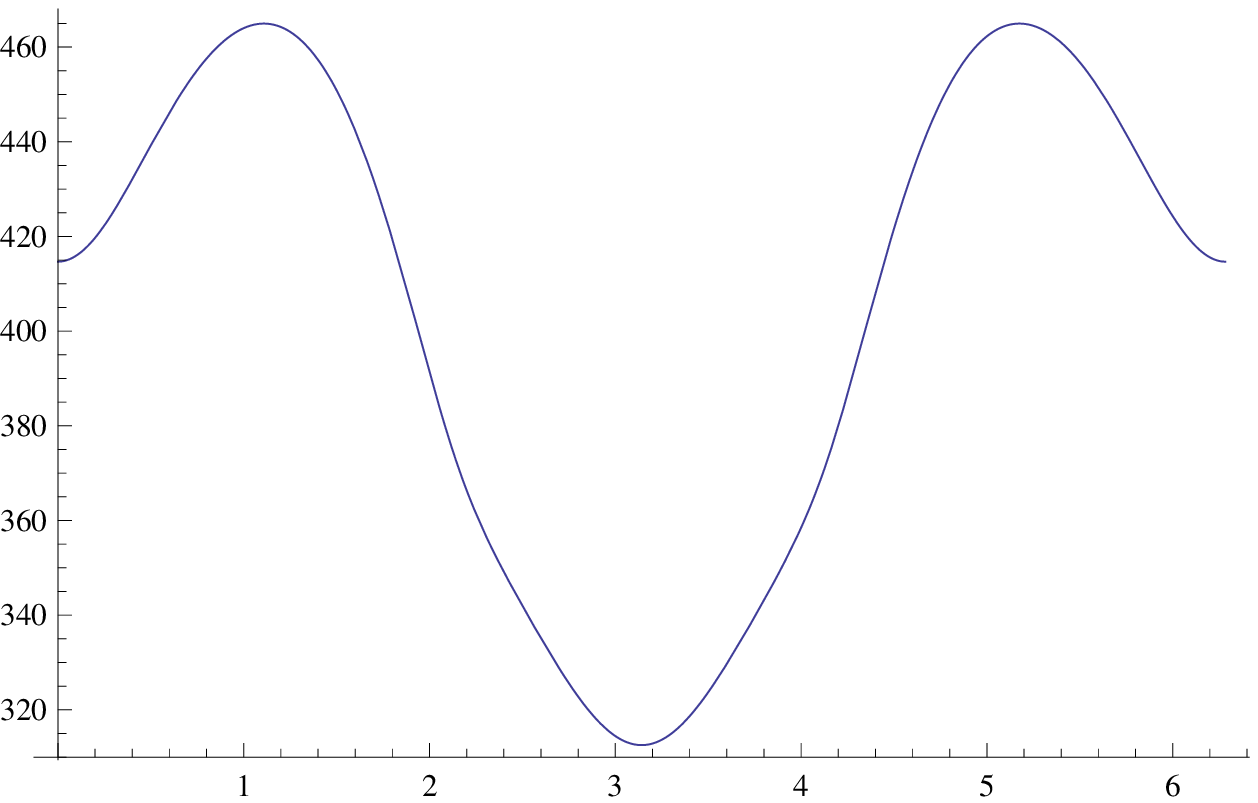}
  \end{center}
  \caption{ $m_{h_2}$  [in GeV] as a function of $\xi$. }
  \label{h2}
\end{figure}

\begin{figure}[htbp]
 \begin{center}
  \includegraphics[width=0.8\textwidth]{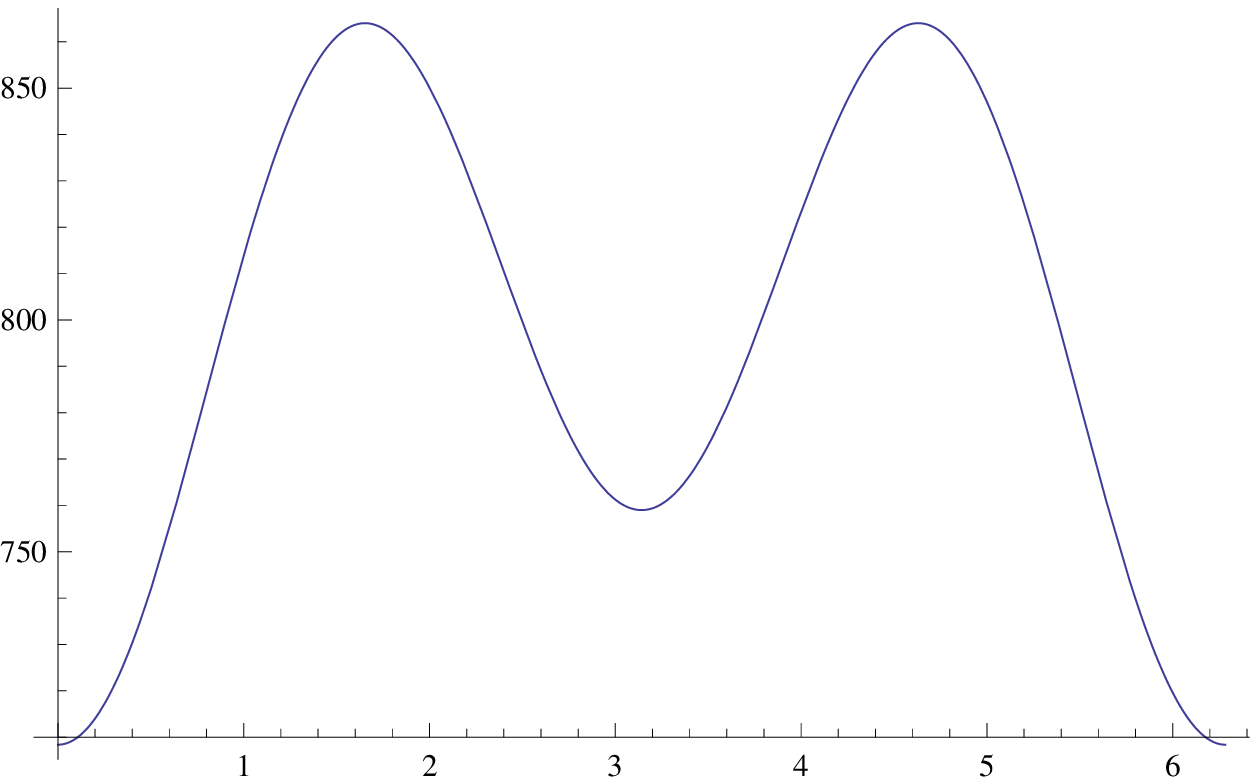}
  \end{center}
  \caption{ $m_{h_3}$ [in GeV] as a function of $\xi$. }
  \label{h3}
\end{figure}

Here matrix $V_n$ can be parameterized by three angles $\theta_i (i=1-3) $ with $0\leq \theta_i \leq \pi/2$. The physical quantities can be chosen as $\left(v, \tan\beta,\xi,
m_{h_1},m_{h_2},  m_{h_3}, \theta_1,\theta_2,\theta_3, m_{H^\pm} \right)$ in accordance with the degree of freedom in Higgs potential. Because the analytical transformation from $\lambda_i$ to Higgs boson masses and the mixing angles are too lengthy to be useful here,
 for our purpose, we choose a set of $\lambda_i $ parameters to illustrate the connection between Higgs boson masses  with $\xi$, especially for $m_{h_1}$.
The benchmark point is chosen as
$$
v=175 GeV, \tan\beta=1, \lambda_7=-2, \lambda_1=6, \lambda_{2,3,4,5,6}=1.
$$
This point satisfies the inequalities (\ref{ieq}), especially the $\xi \rightarrow 0$ is the physical case. In fact there does have parameters in which $\xi=0$ is not allowed. The lightest Higgs boson mass $m_{h_1}$, $m_{h_2}$ and $m_{h_3}$ are depicted as a function of $\xi$ in Figs. \ref{h1}-\ref{h3} respectively. From the figure it is quite clear the dependence of light Higgs approaches massless for the CP conserving
case.
More importantly, besides the lightest Higgs boson, the other neutral Higgs bosons are much heavier. These three figures clearly demonstrate the crucial role of the spontaneous CP violation parameter $\xi$.


\section{Discussion and conclusion}

To summarize, we reveal in this short paper the intimate connection between the spontaneous CP violation and the lightness of the Higgs boson, which was discovered recently by ATLAS and CMS. In the limit of CP conserving case, the lightest Higgs boson degenerates with the massless neutral Goldstone boson. Such connection implies that there should be more Higgs sectors and additional CP violation arising from the complex vacuum. The newly discovered state with mass around 125 GeV is not a pure CP-even scalar as in the standard model. Instead it is a mixing state of CP-even and CP-odd. Besides the lightest neutral Higgs bosons, there are more much heavier neutral Higgs bosons.
We demonstrate this point in the two-Higgs doublet model and which can be true in the more complicated case.
For the two-Higgs doublet model, current experiments are insensitive to neutral Higgs bosons because the couplings are usually severely suppressed by $m_f/m_W$ for light fermions. It is not hard to find allowed parameters. On the contrary the couplings of top quark with Higgs boson are not suppressed. However the huge QCD backgrounds may bury the signal. Such processes deserve further investigations. Actually the full phenomenological investigations are beyond the scope of this paper.
Last but not least, besides the current running colliders, dedicated Higgs factory may be needed in order to precisely measure the detailed information of the lightest neutral Higgs boson. The whole Higgs spectrum and couplings can only be revealed at the next generation colliders.

\section*{Acknowledgment}

I would like to thank Y.N. Mao to cross-check part of the formulas and this work was supported in part by the Natural Science Foundation
 of China (Nos. 11075003 and 11135003).

\bibliographystyle{h-physrev}
\bibliography{reference}

\begin{thebibliography}{10}

\bibitem{:2012gk}
G.~Aad {\em et~al.}, ATLAS Collaboration,
\newblock Phys.Lett. {\bf B716}, 1 (2012), 1207.7214.

\bibitem{:2012gu}
S.~Chatrchyan {\em et~al.}, CMS Collaboration,
\newblock Phys.Lett. {\bf B716}, 30 (2012), 1207.7235.

\bibitem{Sakharov:1967dj}
A.~Sakharov,
\newblock Pisma Zh.Eksp.Teor.Fiz. {\bf 5}, 32 (1967).

\bibitem{Riotto:1999yt}
A.~Riotto and M.~Trodden,
\newblock Ann.Rev.Nucl.Part.Sci. {\bf 49}, 35 (1999), hep-ph/9901362.

\bibitem{Lee:1973iz}
T.~Lee,
\newblock Phys.Rev. {\bf D8}, 1226 (1973).

\bibitem{Lee:1974jb}
T.~Lee,
\newblock Phys.Rept. {\bf 9}, 143 (1974).

\bibitem{Huang:1999xa}
C.-s. Huang and S.-H. Zhu,
\newblock Phys.Rev. {\bf D61}, 015011 (2000), hep-ph/9905463.

\bibitem{Huang:2001me}
C.-S. Huang, W.~Liao, Q.-S. Yan, and S.-H. Zhu,
\newblock Eur.Phys.J. {\bf C25}, 103 (2002), hep-ph/0110147.

\bibitem{Weinberg:1976hu}
S.~Weinberg,
\newblock Phys.Rev.Lett. {\bf 37}, 657 (1976).

\bibitem{Branco:1980sz}
G.~C. Branco,
\newblock Phys.Rev. {\bf D22}, 2901 (1980).

\bibitem{Shizuya:1980jx}
K.~Shizuya and S.~Tye,
\newblock Phys.Rev. {\bf D23}, 1613 (1981).

\end{thebibliography}

\end{document}